\documentclass[11pt]{article}

\usepackage[final]{acl}

\usepackage{times}
\usepackage{latexsym}
\usepackage{amsmath}
\usepackage[most]{tcolorbox} \usepackage{enumitem}
\usepackage[T1]{fontenc}

\usepackage[utf8]{inputenc}

\usepackage{microtype}

\usepackage{inconsolata}

\usepackage{graphicx}

\usepackage{booktabs}
\usepackage{multirow}
\usepackage{enumitem}
\usepackage{pifont}

\usepackage{amssymb}
\usepackage{fvextra}

\usepackage{xcolor}
\newcommand{\highlight}[1]{\textcolor{black}{#1}}

\title{Event-Grounded Question Answering over Long Audio via Structured Retrieval}

\author{
  Kartik Hegde\textsuperscript{1} \quad
  Arvind Krishna Sridhar\textsuperscript{2} \quad
  Naveen Vakada\textsuperscript{1}\thanks{Work done while at Qualcomm Technologies, Inc.} \quad
  Yinyi Guo\textsuperscript{2} \quad
  Erik Visser\textsuperscript{2} \\
  \textsuperscript{1} Qualcomm Technologies, Inc., India \\
  \textsuperscript{2} Qualcomm Technologies, Inc., USA \\
  \texttt{\{karthegd, nvakada, arvisrid, yinyig, evisser\}@qti.qualcomm.com} 
}

\begin{document}
\maketitle
\begin{abstract} Answering natural-language questions over multi-hour audio requires reliable event recognition, temporal grounding, and efficient retrieval. We present LA-RAG (Long Audio Retrieval-Augmented Generation), a structured framework that converts audio into timestamped event records, stores them in an event database, and answers questions using intent-aware retrieval and LLM-based generation. LA-RAG supports two deployment settings: offline grounding mode, in which long recordings are pre-indexed for low-latency querying, and inference-time grounding mode, which performs query-conditioned grounding over shorter, open-ended clips. We evaluate LA-RAG on two synthetic 24-hour benchmarks, Home-IoT and Industrial-IoT, and on CASTELLA-QA and SpotSound-Bench-QA, which are QA versions of two public audio temporal grounding benchmarks that we construct. The results demonstrate effective long-audio question answering with low query-time latency after grounding and indexing. They also reveal a substantial gap between event detection and temporal localization in current large audio-language models. Explicit timestamped grounding consistently improves temporal localization while maintaining competitive detection accuracy. These findings establish structured grounding and retrieval as a practical complement to generative audio-language models for deployment-oriented long-audio understanding. \end{abstract}

\section{Introduction}
Large audio-language models (LALMs)~\cite{yang2024qwen25, yang2025qwen3, xu2025qwen3omni, goel2025af3, ding2025kimiaudiotechnicalreport} perform well on short audio tasks such as captioning, question answering, and instruction following. However, real-world applications often involve continuous audio streams lasting minutes or hours~\cite{chronosaudio, af2}. In industrial monitoring, smart homes, surveillance, and meetings, users ask event-centric questions such as ``When did the alarm sound?'' or ``How many times did the machine stop?'', which require both event recognition and timestamped evidence.

Directly applying LALMs to long audio is difficult: multi-hour recordings exceed practical context and compute limits, and existing models often detect events without accurately localizing their temporal boundaries. Sliding-window inference is a workaround for handling long audio with LALMs \cite{listeningwithtime}, but it is inefficient for long audio QA. Recent long-audio benchmarks primarily target general long-audio understanding or temporal awareness tasks~\cite{chronosaudio, listeningwithtime}, while some relevant resources are not publicly available~\cite{af2, voicegiraffe}. 
\highlight{We therefore convert two publicly available temporal grounding benchmarks, CASTELLA~\cite{castella} and SpotSound-Bench~\cite{spotsound}, into QA format to evaluate event-grounded audio question answering.}

We propose LA-RAG, a structured retrieval-augmented framework that converts audio into timestamped event records using an Audio Grounding Model, stores them in a SQL event store, and retrieves relevant evidence through intent-aware query processing. By decoupling acoustic grounding from language generation, LA-RAG enables scalable, low-latency, and auditable question answering over long audio.

Our contributions are:
\begin{itemize}
    \item We propose \textsc{LA-RAG}, a structured retrieval-augmented framework for long-audio question answering that converts continuous audio streams into timestamped event records and performs intent-aware retrieval over an explicit event store.
    \item We introduce two complementary grounding modes: \emph{offline grounding mode}, which pre-indexes multi-hour audio streams to support efficient question answering, and \emph{inference-time grounding mode}, which performs query-conditioned temporal grounding over short open-ended audio clips.
    \item \highlight{We evaluate LA-RAG on synthetic long-form IoT benchmarks and QA-augmented versions of CASTELLA and SpotSound-Bench, showing that explicit timestamped grounding improves temporal localization and reveals the temporal-localization limitations of current large audio-language models.}
    \item We demonstrate the feasibility of LA-RAG on an edge platform and show that its downstream AQA component runs on-device with a 4B-parameter model.
\end{itemize}

\section{Related Work}
\label{sec:related_work}

Retrieval-augmented generation (RAG) grounds LLM outputs in retrieved evidence~\cite{lewis2020rag,gupta2024rag_survey} and has been extended to audio and video understanding~\cite{chen2025wavrag,luo2024video_rag,tan2025rag_adapter}. Audio question answering and recent audio-language models support acoustic reasoning~\cite{fayek2019daqa,gong2023ltu,ghosh2024gama,goel2025af3,xu2025qwen3omni}, but processing multi-hour audio remains challenging. Recent benchmarks study long-audio understanding~\cite{chronosaudio,listeningwithtime}, while CASTELLA~\cite{castella} and SpotSound-Bench~\cite{spotsound} specifically evaluate temporal sound localization. We convert the latter two into QA benchmarks. LA-RAG instead represents long audio as timestamped event records and uses structured retrieval, related to text-to-SQL approaches~\cite{hong2024text2sql_survey,deng2021structure}.

\begin{figure*}[!ht]
    \centering
    \includegraphics[width=0.7\linewidth]{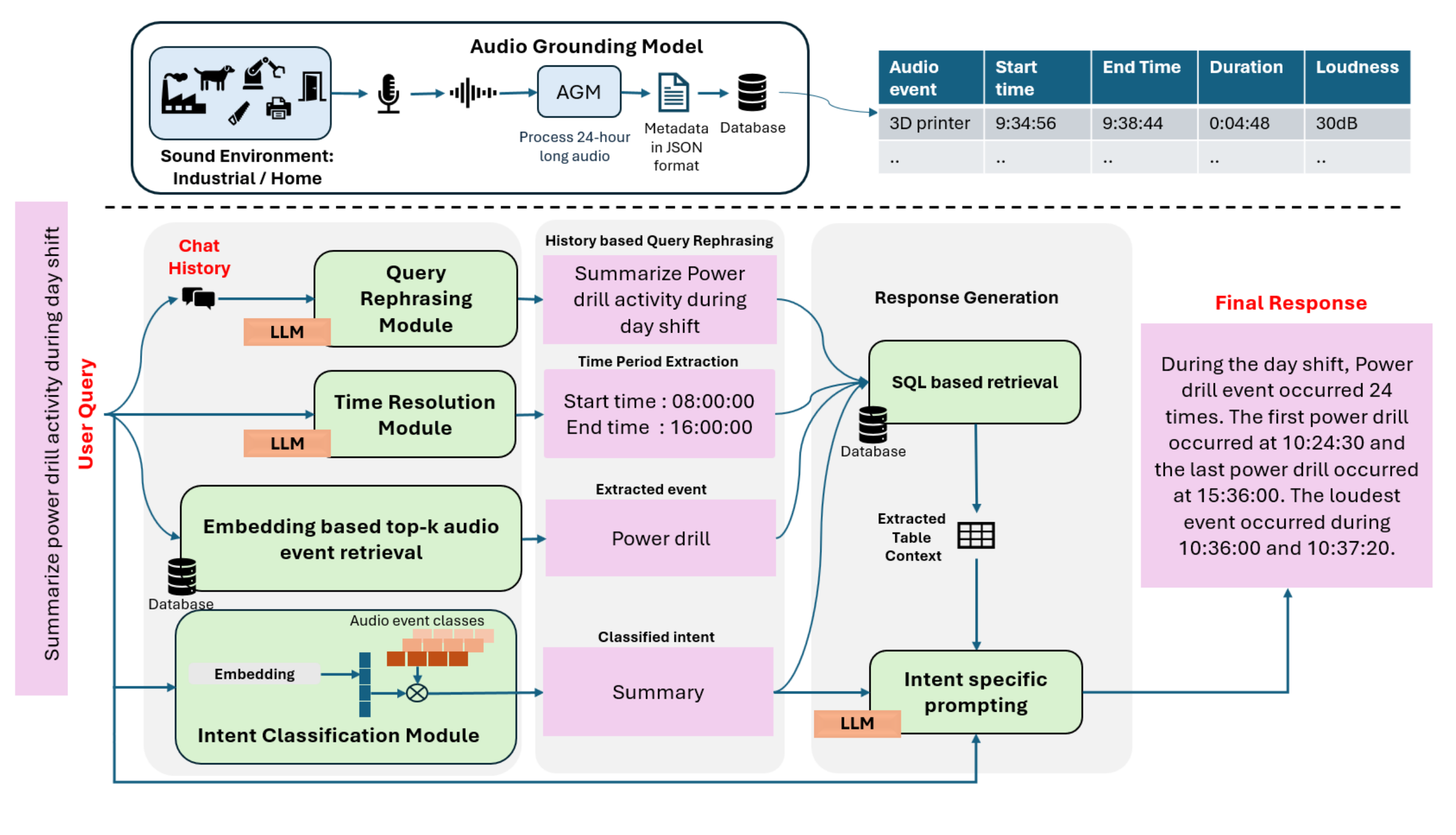}
    \caption{LA-RAG: Proposed method for long-audio question answering via offline audio grounding.}
    \label{fig:main}
\end{figure*}

\section{Methodology}
\label{sec:methodology}

LA-RAG separates acoustic grounding from language reasoning by representing audio as timestamped event records queried through intent-aware retrieval. It supports two modes: \emph{offline grounding}, where long audio is preprocessed into a SQL event store, and \emph{inference-time grounding}, where query-conditioned events are extracted from shorter clips at inference time. Figure~\ref{fig:main} illustrates the pipeline.

\subsection{Audio Grounding Model}
\label{sec:agm}
To preserve the flexibility required for on-device deployment across diverse acoustic environments and application scenarios, we develop an open-vocabulary Sound Event Detection (SED) model based on text-to-audio grounding, comprising approximately 8M parameters. Instead of relying on a fixed label set, the model localizes sound events conditioned on free-form textual queries corresponding to the sounds of interest for a given use case. Our Audio Grounding Model (AGM) is trained following the phrase-level WSTAG framework described in \cite{xu2024towards}, using the AudioCaps \cite{kim2019audiocaps} dataset from which we extract phrase-level supervision. AGM adopts the same architecture and hyperparameters as \cite{xu2024towards}: an audio encoder built from a CRNN with eight convolutional layers followed by a bidirectional GRU (BiGRU), and a text encoder composed of a single word embedding layer with mean pooling. Frame-level grounding scores are computed via the cosine similarity between audio frame embeddings and text tag embeddings, followed by a sigmoid activation. The resulting event records are first serialized in JSON format and then inserted into a SQLite database. The database contains columns for the event name, start time, end time, class confidence, and optional attributes such as loudness.

\subsection{Offline Grounding Mode}
\label{sec:offline-grounding}
Offline grounding mode targets long-form audio for which event metadata can be computed before queries arrive. AGM processes the audio using overlapping windows and stores the detected events as structured SQL records containing event labels, timestamps, confidence scores, and optional attributes. At query time, LA-RAG rephrases context-dependent queries, resolves temporal expressions, and assigns one of four intents: \textit{detection}, \textit{counting}, \textit{summary}, or \textit{anomaly}. It then retrieves relevant events using structured temporal filters and embedding similarity. The retrieved records are inserted into an intent-specific prompt for LLM-based answer generation. Because queries operate on compact event records rather than raw audio, this mode supports efficient and auditable question answering over multi-hour recordings. Appendix~\ref{app:grounding_details} provides the implementation details.

\subsection{Inference-Time Grounding Mode}
\label{sec:inference-time-grounding}

Inference-time grounding mode is used for shorter, unconstrained audio when a precomputed event log is unavailable or unnecessary. After query rephrasing, a rule-based phrase extractor identifies the queried sound event and supplies it to AGM as a text condition. For compound queries, each extracted event phrase is grounded independently. If no phrase can be extracted, AGM evaluates a fallback vocabulary of 75 AudioSet-derived acoustic tags. AGM returns timestamps and confidence scores for the detected query-relevant events. These event records, together with the query and its inferred intent and temporal constraints, are provided to an LLM or to a LALM that also receives the audio. Unlike offline grounding, this mode does not require prior indexing and is not limited to a fixed query vocabulary. It is therefore suitable for event detection and temporal localization in short, open-ended audio clips. Appendix~\ref{app:grounding_details} provides the complete pipeline.

\subsection{Deployment Considerations}
\label{subsec:deployment}

We implemented LA-RAG with offline audio grounding as an on-device prototype. The Audio Grounding Model (AGM) and the downstream audio question answering (AQA) service run as individual microservices on the Qualcomm IQ-9075 EVK~\cite{qualcomm2026iq9075} platform. We use Qwen3-4B-2507 as the on-device language model because it is among the models supported for accelerated inference on the platform's neural processing unit (NPU). This platform delivers up to 100 TOPS of on-device AI performance. Its octa-core Kryo Gen6 CPU, Hexagon Tensor Processors, and industrial operating range support continuous edge inference. The deployment architecture keeps raw audio processing on the edge device, avoiding the need to transmit raw audio to a cloud service. Appendix~\ref{app:ondevice} describes the implementation and reports an on-device evaluation of the AQA component. 


\begin{table*}[t] \centering \small \setlength{\tabcolsep}{6pt} \caption{ \highlight{Results on the long-form IoT benchmarks in offline grounding mode. Detection questions are evaluated using binary accuracy, counting questions using exact-match (EM) accuracy, and summary questions using GPT-4o as the judge on a 1-5 scale. Higher values indicate better performance. Summary responses are evaluated for factual coverage, temporal and causal correctnes and, reasoning coherence. Latency denotes average query-time latency measured on an NVIDIA A100 GPU after audio grounding and database construction. It excludes one-time AGM inference, database construction, and model-loading costs. }} \label{tab:iot_results} \begin{tabular}{llcccc} \toprule \textbf{Dataset} & \textbf{System} & \textbf{Detection Acc. (\%) $\uparrow$} & \textbf{Counting EM (\%) $\uparrow$} & \textbf{Summary (1-5) $\uparrow$} & \textbf{Latency (s) $\downarrow$} \\ \midrule \multirow{3}{*}{Home-IoT} & AGM + RAG & 60.00 & 27.00 & 1.385 & 3.26 \\ & AGM + Text2SQL & 39.33 & 37.00 & 1.220 & 1.34 \\ & \textbf{LA-RAG (Ours)} & \textbf{94.00} & \textbf{75.33} & \textbf{2.805} & \textbf{0.56} \\ \midrule \multirow{3}{*}{Industrial-IoT} & AGM + RAG & 57.00 & 32.00 & 1.465 & 3.66 \\ & AGM + Text2SQL & 46.67 & 34.00 & 1.235 & 1.08 \\ & \textbf{LA-RAG (Ours)} & \textbf{92.00} & \textbf{53.33} & \textbf{2.445} & \textbf{0.44} \\ \bottomrule \end{tabular} \end{table*}

\begin{table*}[t] \centering \small \setlength{\tabcolsep}{5pt} \caption{ Results on CASTELLA-QA and SpotSound-Bench-QA. Detection is evaluated using accuracy. Temporal grounding is evaluated using Recall@1 (R1) at IoU thresholds of 0.3, 0.5, and 0.7, and mean IoU (mIoU). Latency is reported separately for detection and temporal grounding. Input indicates what the answering model receives: the raw audio only, the raw audio and AGM event records, or the AGM event records only. } \label{tab:grounding_results} \begin{tabular}{@{}ll c cccc cc@{}} \toprule & & \textbf{Detection} & \multicolumn{4}{c}{\textbf{Temporal Grounding (\%)}} & \multicolumn{2}{c}{\textbf{Latency (s) $\downarrow$}} \\ \cmidrule(lr){3-3}\cmidrule(lr){4-7}\cmidrule(l){8-9} \textbf{Model} & \textbf{Input} & \textbf{Acc. (\%)} & \textbf{R1@0.3} & \textbf{R1@0.5} & \textbf{R1@0.7} & \textbf{mIoU} & \textbf{Det.} & \textbf{Temp.} \\ \midrule \multicolumn{9}{@{}l}{\textit{CASTELLA-QA}} \\ \midrule Qwen2.5-Omni-7B & Audio only & \textbf{88.00} & 12.50 & 7.50 & 3.50 & 9.00 & 12.28 & 28.35 \\ & Audio + AGM & 80.00 & \textbf{32.50} & \textbf{24.00} & \textbf{18.50} & \textbf{25.40} & 10.20 & 27.87 \\ & AGM only & 64.50 & 31.00 & 21.50 & 16.50 & 23.13 & \textbf{0.35} & \textbf{0.94} \\ \cmidrule(l){2-9} Audio Flamingo 3 & Audio only & 81.00 & 2.50 & 1.00 & 1.00 & 2.18 & 0.98 & 1.46 \\ & Audio + AGM & \textbf{85.50} & \textbf{34.50} & \textbf{25.00} & \textbf{19.00} & 25.83 & 0.97 & 2.31 \\ & AGM only & 78.50 & 33.50 & \textbf{25.00} & 18.50 & \textbf{26.10} & \textbf{0.06} & \textbf{0.42} \\ \midrule \multicolumn{9}{@{}l}{\textit{SpotSound-Bench-QA}} \\ \midrule Qwen2.5-Omni-7B & Audio only & \textbf{91.00} & 24.00 & 14.00 & 7.00 & 16.36 & 1.52 & 3.32 \\ & Audio + AGM & 88.00 & \textbf{56.50} & \textbf{44.00} & \textbf{33.50} & \textbf{43.85} & 1.48 & 3.76 \\ & AGM only & 83.50 & 53.00 & 42.50 & 33.00 & 41.89 & \textbf{0.37} & \textbf{0.91} \\ \cmidrule(l){2-9} Audio Flamingo 3 & Audio only & \textbf{89.50} & 6.00 & 1.00 & 0.50 & 4.35 & 0.24 & 0.82 \\ & Audio + AGM & 88.50 & \textbf{53.00} & 42.50 & 32.50 & \textbf{41.96} & 0.23 & 0.82 \\ & AGM only & 51.50 & \textbf{53.00} & \textbf{43.00} & \textbf{33.00} & 41.84 & \textbf{0.05} & \textbf{0.45} \\ \bottomrule \end{tabular} \end{table*}

\section{Experiments}
\label{sec:experiments}

\subsection{Datasets and Evaluation Setup} \label{sec:datasets} \paragraph{Offline Grounding Mode} To the best of our knowledge, no public benchmark provides event-grounded QA over multi-hour audio with exact timestamp ground truth. We therefore construct two controlled long-form IoT benchmarks \highlight{to evaluate multi-hour retrieval and structured temporal reasoning in LA-RAG}. Each benchmark contains a synthetically constructed 24-hour recording \highlight{representing a home-monitoring (Home-IoT) or factory-floor (Industrial-IoT) environment}. The recordings are generated by inserting short labeled foreground events at known timestamps, thereby providing construction annotations with exact event labels and temporal boundaries. These construction annotations are independent of the AGM predictions used during evaluation. The foreground events are sourced from proprietary licensed sound-effect libraries. They are inserted without temporal overlap at an average rate of approximately three occurrences per class per hour. The event classes are listed in Appendix~\ref{app:classes}. Acoustic events are restricted to these predefined class lists, and the same bounded application ontology is provided to the downstream QA system. The benchmarks therefore evaluate structured reasoning over a fixed monitoring vocabulary rather than open-vocabulary event discovery. Background audio consists of industrial machine and white-noise recordings for Industrial-IoT and low-hum ambience for Home-IoT, sourced from Freesound~\cite{fonseca2021fsd50k}. The foreground events are mixed directly with the background audio at a global SNR of 6~dB. No spatialization, reverberation, distance variation, microphone simulation, codec effects, or additional channel processing is applied. The resulting audio is normalized to the Apple loudness target of $-16$~LUFS. More details on the construction of the QA pairs are provided in Appendix~\ref{app:qagen_offline}.

\noindent\textbf{Evaluation:}
For both IoT benchmarks, detection QA is evaluated as binary classification over the presence or absence of a queried sound event. Counting QA is evaluated using exact match between predicted and ground-truth event counts. Summary QA is evaluated using an LLM-based scoring mechanism. \highlight{We report accuracy for detection and exact match accuracy for counting, GPT based summary evaluation score on a scale of 1-5 and query-time latency after audio grounding and database construction.}

\noindent\textbf{Human validation of summary evaluation:} We validate the GPT-4o summary evaluation using five human raters and a stratified sample of LA-RAG outputs. Across 45 questions with sufficient human ratings, GPT-4o scores show strong agreement with mean human scores (Spearman's $\rho=0.71$, 95\% bootstrap CI: $[0.54,0.84]$), with stronger agreement for event-specific summaries ($\rho=0.80$). These results support GPT-4o scoring as an informative proxy for evaluating LA-RAG summaries.

\paragraph{Inference-Time Grounding Mode} \highlight{We use two public audio temporal grounding benchmarks. CASTELLA~\cite{castella} is a real-world audio moment-retrieval dataset containing approximately five-minute audio clips with manually annotated local captions and event timestamps. Each audio file contains one or more annotated moments, with a natural-language caption and corresponding start and end times. SpotSound-Bench~\cite{spotsound} is a needle-in-a-haystack benchmark for audio temporal grounding, in which short in-the-wild target sound events are embedded within long, unstructured recordings and annotated with their start and end times. We augment the test splits of both datasets with QA pairs derived from these annotations, and refer to the resulting benchmarks as CASTELLA-QA and SpotSound-Bench-QA. Both benchmarks contain detection and temporal grounding questions, which evaluate event recognition and temporal localization, respectively.\\}
\noindent\textbf{Detection QA:} We construct yes/no questions of the form \textit{``Is \{local\_caption\} present in the audio?''} The positive set contains 100 event-present questions derived from local captions associated with the target clips. For each target clip, candidate negative events are sampled from the pool of captioned CASTELLA events, subject to the condition that the selected event caption is absent from that clip's annotations. The negative set is randomly capped at 100 questions to balance the 100 positive questions. These negatives are annotation-absent rather than acoustically verified negatives; therefore, an event not mentioned in a clip's annotations may still be acoustically present. \highlight{For SpotSound-Bench-QA, we follow the same procedure, using the annotated target-event phrases in place of local captions and sampling negatives from events absent from each clip's annotations.}\\
\noindent\textbf{Temporal Grounding QA:} We construct 200 questions of the form \textit{``When does \{local\_caption\} occur in the audio?''} Each question is derived from one annotated local caption with corresponding start and end boundaries. The reference answer is the temporal segment associated with that caption in the CASTELLA annotations. \highlight{The SpotSound-Bench-QA temporal questions are constructed in the same way from the annotated target-event boundaries from SpotSound-Bench.}



\noindent\textbf{Evaluation:}
\highlight{We report \textit{Accuracy} for detection questions. Temporal Grounding is evaluated using Recall@1 (R1) at Intersection-over-Union (IoU) thresholds of 0.3, 0.5, and 0.7, together with mean IoU (mIoU). We also report average latency separately for detection and temporal grounding questions. The average AGM inference latency per query is 54\,ms on CASTELLA-QA and 7\,ms on SpotSound-Bench-QA.}

\subsection{Baselines and Compared Systems}
\label{sec:baselines}

\paragraph{Offline Grounding Mode Baselines.}
We compare LA-RAG with two structured retrieval baselines on the long-form IoT benchmarks:

\begin{itemize}[leftmargin=*, noitemsep]
    \item \textbf{AGM + RAG:} AGM-extracted event logs combined with standard embedding-based retrieval and LLM generation, without intent classification or explicit time resolution.

    \item \textbf{AGM + Text2SQL:} AGM-extracted event logs retrieved through natural-language-to-SQL generation, without intent-aware prompting.
\end{itemize}

\highlight{Current LALMs cannot process 24-hour recordings within their input limits, and no public system targets event-grounded QA at this scale; we therefore compare retrieval strategies over the same event database, and evaluate external LALMs in the inference-time setting (Table~\ref{tab:grounding_results}).}

All three offline-grounding systems evaluated in Table \ref{tab:iot_results} use Phi-3.5-MoE~\cite{phi3technicalreport2024} as the language-model backbone. Additional LLM scaling results are provided in Appendix~\ref{app:IIoT_model_ablation}.

\paragraph{Inference-Time Grounding Mode Baselines.}
\highlight{On CASTELLA-QA and SpotSound-Bench-QA, we evaluate two LALMs, Qwen2.5-Omni-7B and Audio Flamingo 3, under three conditions: (i) \textbf{no grounding}, where the model receives the audio clip and the question; (ii) \textbf{AGM + audio}, where the model also receives the AGM timestamped event records (event\_start and event\_end); and (iii) \textbf{AGM only}, where the model receives the AGM event records and the question but no raw audio.}

\section{Results and Discussion}

Table~\ref{tab:iot_results} reports results on the long-form IoT benchmarks in offline grounding mode. LA-RAG outperforms AGM+RAG and AGM+Text2SQL on all three question types on both benchmarks. On Home-IoT, LA-RAG achieves 94.00\% detection accuracy and 75.33\% counting EM, compared with 60.00\% and 37.00\% for the strongest baseline on each metric. On Industrial-IoT, it achieves 92.00\% detection accuracy and 53.33\% counting EM, compared with 57.00\% and 34.00\%. LA-RAG also obtains the highest summary scores (2.805 and 2.445 on a 1-5 scale). However, summary questions and Industrial-IoT counting remain challenging, which indicates room for improvement on multi-event aggregation. LA-RAG also achieves the lowest query-time latency (see Table~\ref{tab:iot_results} for scope). On Home-IoT, it reduces latency to 0.56\,s, compared with 3.26\,s for AGM+RAG and 1.34\,s for AGM+Text2SQL. On Industrial-IoT, it achieves 0.44\,s, an 8.32$\times$ speedup over AGM+RAG. We observe that the embedding-based RAG baseline can retrieve incorrect evidence for complex or temporally constrained queries, while Text2SQL can fail when the model generates an incorrect SQL query, resulting in no valid answer. These gains are consistent with the benefits of intent-aware retrieval and explicit time resolution, which help mitigate these failure modes. Removing time resolution reduces Industrial-IoT detection accuracy from 92.00\% to 77.33\% and counting EM from 53.33\% to 17.00\% (Appendix~\ref{app:time_resolution}). Because the benchmarks use controlled acoustic conditions, these results mainly reflect long-duration retrieval and temporal reasoning rather than acoustic robustness.

\highlight{Table~\ref{tab:grounding_results} evaluates inference-time grounding on CASTELLA-QA and SpotSound-Bench-QA. Ungrounded LALMs achieve strong detection accuracy but poor temporal localization, indicating that they can detect event presence but struggle to estimate event boundaries. Adding AGM consistently improves temporal localization for both models on both benchmarks. On CASTELLA-QA, mIoU increases from 9.00\% to 25.40\% for Qwen2.5-Omni and from 2.18\% to 25.83\% for Audio Flamingo 3. On SpotSound-Bench-QA, mIoU increases from 16.36\% to 43.85\% and from 4.35\% to 41.96\%, respectively. Detection accuracy remains competitive: AGM lowers detection accuracy for Qwen2.5-Omni (e.g., 88.00\% to 80.00\% on CASTELLA-QA) but raises it for Audio Flamingo 3 on CASTELLA-QA (81.00\% to 85.50\%). When the raw audio is also provided, adding AGM has a small effect on latency in most cases. Removing direct audio input (AGM only) preserves most of the temporal-grounding gain. R1@0.5 stays within 2.5 points of the AGM + audio setting in all cases, and Audio Flamingo 3 matches or slightly exceeds it. This setting also substantially reduces latency; for example, Qwen2.5-Omni temporal grounding latency on CASTELLA-QA falls from 27.87\,s to 0.94\,s. However, detection accuracy decreases without direct audio, most notably for Audio Flamingo 3 on SpotSound-Bench-QA (88.50\% to 51.50\%). In this setting, detection depends fully on AGM outputs, so grounding errors propagate directly to the answer. These results indicate that explicit timestamped grounding is the main driver of temporal localization, while direct audio access mainly supports event detection. Additional comparisons between language-only and multimodal models are provided in Appendix~\ref{app:larag_inference_time}, and replacing AGM with a publicly available grounding model is studied in Appendix~\ref{app:openflamvsagm}.} \highlight{One-time AGM inference takes about 0.4\,ms per second of audio on a single A100 GPU (about 35\,s for a 24-hour recording).}

\section{Conclusion}

We introduced LA-RAG, a retrieval-augmented framework for event-grounded question answering over long audio. By converting audio into timestamped event records and retrieving evidence through intent-aware query processing, LA-RAG decouples acoustic grounding from language generation, enabling efficient reasoning over multi-hour recordings. LA-RAG consistently outperforms AGM+RAG and AGM+Text2SQL across detection, counting, and summary tasks with minimal latency. \highlight{On CASTELLA-QA and SpotSound-Bench-QA, AGM-based grounding consistently improves temporal localization, confirming that explicit timestamped evidence is critical for temporal audio QA. Grounding-only inference without direct audio input preserves most of this gain at substantially lower latency, although detection accuracy decreases. These results suggest that structured audio retrieval is a practical and scalable complement to generative audio-language models.}

\section{Limitations} \label{sec:limitations} Our experiments show that the LA-RAG pipeline is effective and efficient for long-audio question answering. For hour-long audio, larger and stronger language models could further improve accuracy. However, on-device deployment requires models with a low memory footprint. \highlight{On real-world audio clips, AGM-based grounding consistently improves temporal localization on CASTELLA-QA and SpotSound-Bench-QA, but the absolute temporal-localization scores remain moderate, especially on CASTELLA-QA.} Fine-tuning the grounding model for the target domain could improve these scores. The end results depend on the performance of the audio grounding model. This dependency is most visible in grounding-only inference without direct audio input: latency drops substantially, but detection accuracy decreases because grounding errors propagate directly to the answer. Counting and summary queries can also be difficult, because they require precise information about every event in the audio. When a query is generic, phrase extraction returns no events. The entire vocabulary is then passed to the grounding model as candidate events, which can increase false alarms. Better tuning and more capable language or multimodal models could address these limitations in future work.

\bibliography{custom}
\newpage

\appendix
\label{sec:appendix}

\clearpage

\section{Audio classes for Home-IoT and Industrial IoT datasets}
\label{app:classes}
The home set covers \emph{alarms, sirens, door\_bell, door\_knock, glass\_breaking, car\_crash, door\_close-open, baby\_cry, gun\_shot, cat, car\_honk, snoring, dog\_bark}; the industrial set includes \emph{tools clanking, hand saw, hand file, workers talking, footsteps, arc welder, diesel forklift, power hand drill, stamping machine, walkie talkie, warning buzzer, factory whistle}

\section{AGM Evaluation} \label{sec:agm_appendix} Table~\ref{tab:sed_leaderboard} reports performance on AudioMarathon Sound Event Detection (SED) task~\cite{he2025audiomarathon}. Our AGM was evaluated following the same task and evaluation protocol, achieving an F1 score of \textbf{74.8}. The results for the other models are reported from~\cite{he2025audiomarathon}. Under this shared protocol, AGM achieves the second-highest F1 score among the listed systems.

\begin{table}[h] \centering \caption{\highlight{Sound Event Detection (SED) results on the AudioMarathon data.} The result marked with $^{*}$ was obtained using our AGM; all other results are reported from~\cite{he2025audiomarathon}.} \label{tab:sed_leaderboard} \resizebox{0.4\textwidth}{!}{ \begin{tabular}{c l c} \hline \textbf{Rank} & \textbf{Model} & \textbf{SED (F1)} \\ \hline 1 & Qwen2.5-Omni-7B & 78.4 \\ \textbf{2} & \textbf{AGM (Ours)}$^{*}$ & \textbf{74.8} \\ 3 & Voxtral-Mini-3B-2507 & 71.0 \\ 4 & Qwen2.5-Omni-3B & 70.2 \\ 5 & Audio-Flamingo-3 & 59.5 \\ 6 & Phi-4-Multimodal & 55.1 \\ 7 & Aero-1-Audio & 55.0 \\ 8 & Gemma-3n-E2B-it & 50.2 \\ 8 & Gemma-3n-E4B-it & 50.2 \\ 10 & Baichuan-Omni-1.5 & 45.7 \\ 11 & Audio-Flamingo-2 & 27.1 \\ \hline \end{tabular} } \end{table}


\begin{table}[h]
\centering
\small
\caption{Encoder ablation on CASTELLA in inference-time grounding mode. Detection accuracy and Temporal Precision (Tem. P), Temporal Recall (Tem. R) and Temporal F1 (Tem.F1) scores are tabulated.}
\label{tab:encoder_ablation}
\setlength{\tabcolsep}{3pt}
\resizebox{\columnwidth}{!}{%
\begin{tabular}{lcccc}
\toprule
\textbf{Grounding model} & \textbf{Det. Acc.} & \textbf{Tem. P} & \textbf{Tem. R} & \textbf{Tem. F1} \\
\midrule
OpenFLAM   & 60.00 & 9.85  & 11.93 & 10.79  \\
AGM (Ours) & \textbf{74.50} & \textbf{21.09} & \textbf{18.96} & \textbf{19.97} \\
\bottomrule
\end{tabular}%
}
\end{table}

\section{Backbone Comparison in LA-RAG with Inference-Time Grounding}
\label{app:larag_inference_time}

This appendix examines whether replacing the text-only backbone of LA-RAG with an audio-language model improves performance. Table~\ref{tab:castella_larag} compares Qwen2.5-7B-Instruct (text only) with Qwen2.5-Omni-7B (text + audio) on CASTELLA in inference-time grounding mode. Both configurations use the full LA-RAG pipeline, including intent classification and time resolution, and the same AGM event records. The Omni model also receives the raw audio. Unlike Table~\ref{tab:grounding_results}, which evaluates the models without the LA-RAG pipeline, this experiment isolates the effect of the backbone model. Access to the raw audio improves detection accuracy from 74.50\% to 84.00\%. Temporal grounding remains comparable: both backbones obtain the same R1@0.3 and R1@0.5, and the Omni model gives only small gains at R1@0.1 (+4.00), R1@0.7 (+1.00), and mIoU (+0.65). This indicates that the AGM event records already provide most of the timing information. However, the Omni model increases average latency. Therefore, the audio-language backbone is preferable when detection accuracy is the priority and latency is less constrained. Otherwise, the text-only backbone gives similar temporal grounding at a much lower cost. All latencies in this experiment are measured on two NVIDIA A100 GPUs.

\begin{table*}[!ht] \centering \caption{Comparison of LA-RAG with offline grounding mode for Industrial IoT across LLMs.} \label{tab:iiot_ablation} \resizebox{\textwidth}{!}{ \begin{tabular}{l c c c c} \hline \textbf{Model Name} & \textbf{\# Active Params} & \textbf{Detection (\%)} & \textbf{Counting EM (\%)} & \textbf{Summary (1-5)} \\ \hline Phi-3-medium \cite{phi3technicalreport2024} & 14B & 93.33 & 47.00 & 2.430 \\ Minitron \cite{muralidharan2024compact} & 8B & 92.67 & 44.33 & 2.135 \\ Qwen3-8B \cite{yang2025qwen3} & 8B & 86.33 & 24.00 & 1.845 \\ Llama-3.1 \cite{grattafiori2024llama} & 8B & 49.67 & 19.67 & 1.315 \\ \hline Qwen2.5 \cite{yang2024qwen25} & 7B & 88.67 & 49.67 & 1.925 \\ Phi-3.5-MoE \cite{phi3technicalreport2024} & 6.6B (42B) & 92.00 & 53.33 & 2.445 \\ Phi-4-mini-instruct \cite{abouelenin2025phi} & 3.8B & 93.67 & 40.67 & 2.190 \\ \hline Llama-3.2 \cite{grattafiori2024llama} & 3B & 41.67 & 15.67 & 1.420 \\ Qwen2.5 \cite{yang2024qwen25} & 0.5B & 64.33 & 7.00 & 1.910 \\ \hline \end{tabular} } \end{table*}

\begin{table*}[t] \centering \small \setlength{\tabcolsep}{5pt} \caption{Backbone comparison within the full LA-RAG pipeline on CASTELLA in inference-time grounding mode. Both rows use AGM event records; the audio-language model also receives the raw audio. Detection is evaluated using overall accuracy, and temporal grounding using R1 at IoU thresholds 0.1, 0.3, 0.5, and 0.7 and mIoU. Latency is reported separately for detection and temporal grounding queries.} \label{tab:castella_larag} \begin{tabular}{lcccccccc} \toprule & \textbf{Detection} & \multicolumn{5}{c}{\textbf{Temporal Grounding}} & \multicolumn{2}{c}{\textbf{Avg.\ Latency (s)} $\downarrow$} \\ \cmidrule(lr){2-2}\cmidrule(lr){3-7}\cmidrule(lr){8-9} \textbf{Backbone (7B)} & \textbf{Overall Acc.} & \textbf{R1@0.1} & \textbf{R1@0.3} & \textbf{R1@0.5} & \textbf{R1@0.7} & \textbf{mIoU} & \textbf{Det.} & \textbf{Temp.} \\ \midrule Qwen2.5-7B-Instruct (text) & 74.50 & 46.00 & 34.00 & 25.00 & 12.00 & 25.38 & \textbf{0.36} & \textbf{0.79} \\ Qwen2.5-Omni-7B (text + audio) & \textbf{84.00} & \textbf{50.00} & 34.00 & 25.00 & \textbf{13.00} & \textbf{26.03} & 2.68 & 42.42 \\ \bottomrule \end{tabular} \end{table*}

\section{AGM Encoder Ablation}
\label{app:openflamvsagm}
Table~\ref{tab:encoder_ablation} compares LA-RAG performance 
using AGM versus publicly available OpenFLAM~\cite{flam2025} as the grounding 
encoder on the CASTELLA benchmark. \highlight{Results demonstrate that LA-RAG can operate with an alternative publicly available grounding model, while downstream performance remains sensitive to grounding quality.}


\section{Comparison Across Model Scales on the Industrial-IoT Dataset} \label{app:IIoT_model_ablation} Table~\ref{tab:iiot_ablation} compares LLMs from 0.5B to 14B active parameters as the response generator in LA-RAG with offline grounding mode. \texttt{Phi-3.5-MoE} (6.6B active, 42B total), which is the default model in Table~\ref{tab:iot_results}, achieves the highest counting EM (53.33\%) and summary score (2.445). \texttt{Phi-3-medium} (14B) is close on summary (2.430) but lower on counting (47.00\%). \texttt{Phi-4-mini-instruct} (3.8B) achieves the highest detection accuracy (93.67\%) despite its smaller size. \highlight{We use \texttt{Phi-3.5-MoE} as the default because it gives the best balance across tasks: it leads on counting and summary, remains within 1.7 points of the best detection accuracy, and uses fewer active parameters than \texttt{Phi-3-medium}.} Model size alone does not determine performance. Among 8B models, detection accuracy ranges from 49.67\% (\texttt{Llama-3.1}) to 92.67\% (\texttt{Minitron}), and counting EM ranges from 19.67\% to 44.33\%. Counting is the most model-sensitive task (7.00\%--53.33\%), while most models above 3B exceed 86\% detection accuracy. Small models degrade clearly on counting: \texttt{Llama-3.2-3B} and \texttt{Qwen2.5-0.5B} reach only 15.67\% and 7.00\% counting EM. These results indicate that aggregating structured event records requires both adequate model capacity and strong instruction-following ability, and that model family matters as much as parameter count.

\section{Time-Resolution Ablation} \label{app:time_resolution} We evaluate the contribution of explicit time resolution on the Industrial-IoT benchmark in offline grounding mode. The ablated system replaces the query-derived time interval with the full-day interval (00:00--24:00). All other LA-RAG components and the Phi-3.5-MoE backbone are unchanged. Each task contains 300 questions. Unresolved or malformed responses are counted as incorrect. Removing time resolution reduces detection accuracy from 92.00\% to 77.33\% and counting exact match from 53.33\% to 17.00\% (Table~\ref{tab:time_resolution_ablation}). Counting degrades more because, without the requested time window, the system returns the event count over the full recording instead of the count within the queried interval. 

\begin{table}[t] \centering \small \setlength{\tabcolsep}{4pt} \caption{Time-resolution ablation on the Industrial-IoT benchmark in offline grounding mode. Each task contains 300 questions.} \label{tab:time_resolution_ablation} \begin{tabular}{lcc} \toprule \textbf{Metric} & \textbf{LA-RAG} & \textbf{w/o time res.} \\ \midrule Detection accuracy (\%) & \textbf{92.00} & 77.33 \\ Counting EM (\%) & \textbf{53.33} & 17.00 \\ \bottomrule \end{tabular} \end{table}


\section{QA Generation for the IoT Benchmarks} \label{app:qagen_offline} 
We generate evaluation question-answer (QA) pairs using a deterministic procedure based on the synthetic benchmark construction metadata. Each benchmark recording is constructed from a six-hour audio sequence that is repeated four times to produce a 24-hour recording. The foreground sound events and their insertion timestamps are predefined during construction. Their event labels, start times, and end times are recorded as construction annotations. These construction annotations serve as the ground truth for QA generation. The Audio Grounding Model (AGM) is not used to generate, modify, or validate the ground-truth annotations or reference answers. AGM is applied only after benchmark construction as part of the system under evaluation. For implementation consistency, the construction annotations and AGM predictions use the same event-record schema, including fields such as event label, start time, and end time.

We construct three primary question categories: detection questions for binary event presence or absence, counting questions to determine the number of event occurrences, and summary questions for temporal and statistical aggregation over a recording. Answers containing temporal information are normalized to the HH:MM:SS format. To provide varied query surface forms, the questions include 12-hour and 24-hour time expressions, shift-based references, before/after constraints, duration-based phrasing, and relative or segment-based intervals. The dataset includes positive questions for events present in the recording and negative questions involving absent events or unrelated event categories. We also introduce synonym-based variations by replacing canonical event labels with contextually equivalent terms, testing semantic matching beyond exact label lookup. All question templates were manually reviewed for correctness and clarity. The resulting QA pairs provide controlled and reproducible distributions across question categories and temporal expressions, supporting systematic evaluation of event-grounded long-audio question answering.


\section{On-Device Deployment} \label{app:ondevice} 
\begin{figure*}[!ht]
    \centering
    \includegraphics[width=0.8\linewidth]{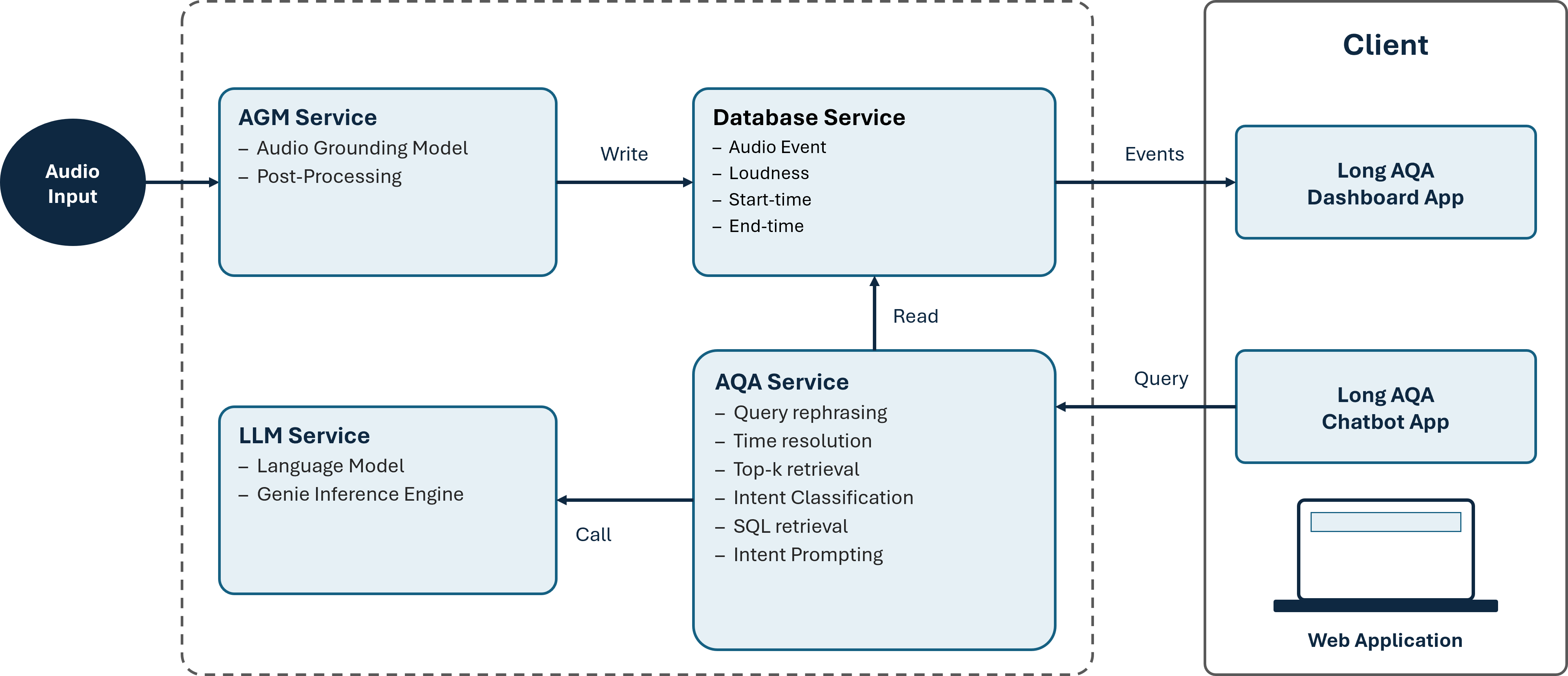} 
    \caption{On-device deployment pipeline}
    \label{fig:ondevice}
\end{figure*}

\begin{figure*}[!ht]
    \centering
    \includegraphics[width=0.8\linewidth]{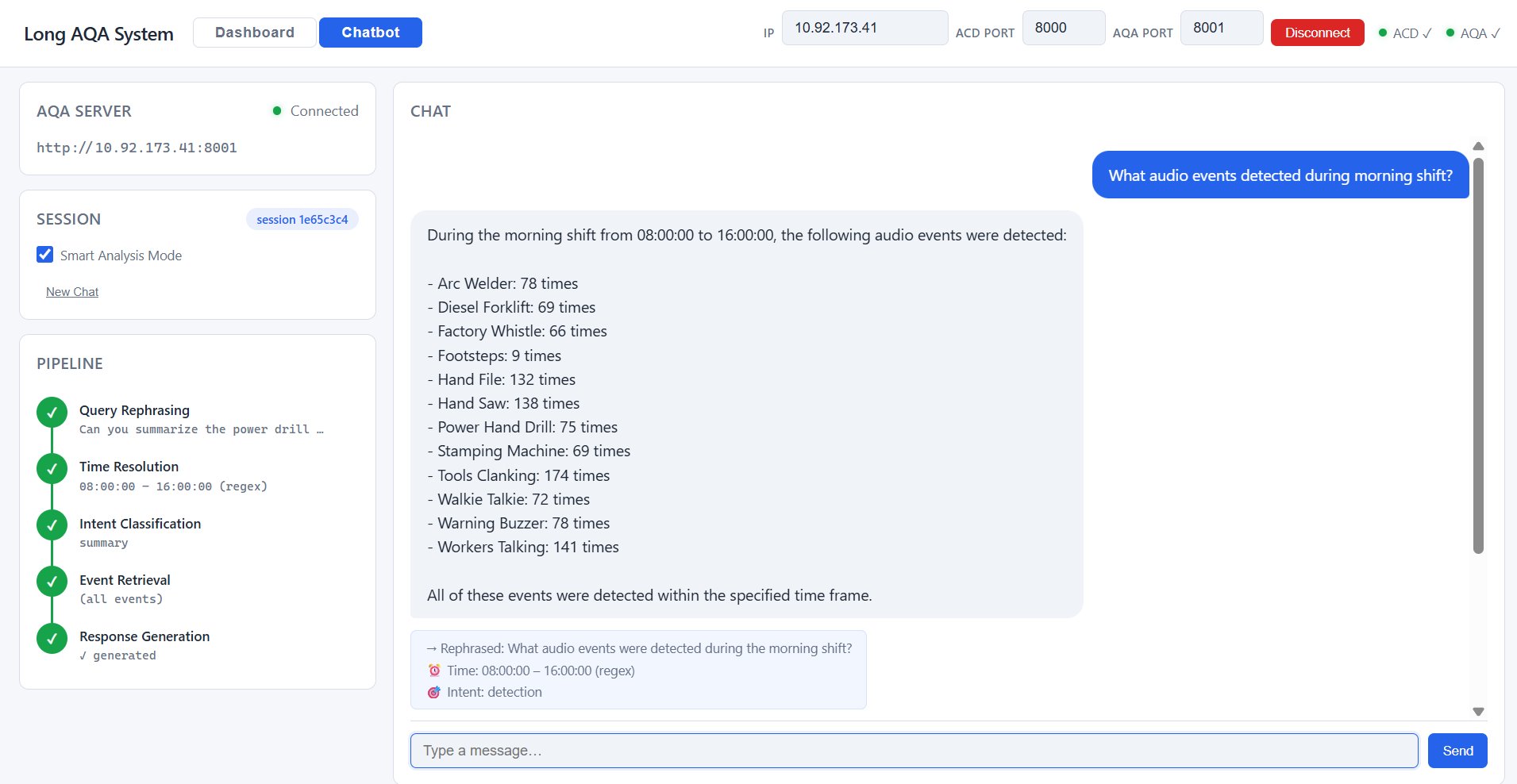} 
    \caption{User Interface of Long Audio QA Chatbot}
    \label{fig:new_ui}
\end{figure*}

We demonstrate the deployment feasibility of LA-RAG using a distributed, hardware-aware microservice architecture on the Qualcomm IQ-9075 EVK, paired with client devices that provide the user-facing interfaces shown in Fig.~\ref{fig:ondevice}. The platform provides heterogeneous compute resources, including CPUs, GPUs, and a neural processing unit (NPU), which support continuous audio processing and on-device language-model inference. This deployment demonstrates that the components of LA-RAG can be integrated and executed in an edge environment without requiring cloud-based audio processing. Incoming audio is processed by the Audio Grounding Model (AGM) described in Section~\ref{sec:agm}. AGM is the same lightweight CRNN-based model used in our experiments, comprising an eight-layer convolutional audio encoder followed by a bidirectional GRU and a word-embedding text encoder. Deployment-stage operations, including temporal smoothing, confidence calibration, and event consolidation, are applied to the frame-level grounding outputs. The resulting event records, including event labels, confidence scores, timestamps, and available attributes such as loudness, are stored in an SQL event database. This event-centric representation supports retrieval over long recordings without repeatedly providing the raw audio to the language model. The Audio Question Answering (AQA) service performs the downstream reasoning. Natural-language queries submitted through the dashboard or chatbot interface are processed through query rephrasing and intent classification. Depending on the inferred intent, the service performs time resolution, event matching, structured temporal filtering, and SQL-based retrieval before constructing an intent-specific prompt for the language model. Language-model inference is performed on-device using a hardware-optimized runtime. The instruction-tuned language model is compiled and quantized into an execution-efficient format compatible with the available acceleration hardware. Because LA-RAG retrieves timestamped event records rather than passing the complete raw audio to the language model, the downstream reasoning stage can operate over long recordings without requiring the complete audio sequence to fit within the language model context. Grounded events can be streamed to visualization dashboards, while interactive question answering is provided through chatbot interfaces, as illustrated in Fig.~\ref{fig:new_ui}. The separation of continuous grounding, structured storage, and intent-aware reasoning provides a modular architecture suitable for edge deployment. To quantitatively evaluate the on-device AQA component, we additionally evaluate the downstream reasoning pipeline on SpotSound-Bench-QA using Qwen3-4B-2507. The evaluation contains 400 questions: 100 positive detection questions, 100 negative detection questions, and 200 temporal grounding questions. 

\begin{table}[t] \centering \small \setlength{\tabcolsep}{4pt} \caption{ On-device evaluation of the downstream AQA component using Qwen3-4B-2507 on SpotSound-Bench-QA. The evaluation uses 100 positive detection, 100 negative detection, and 200 temporal grounding questions.} \label{tab:ondevice_results} \begin{tabular}{lcc} \toprule \textbf{Task} & \textbf{Metric} & \textbf{Result} \\ \midrule \multirow{4}{*}{Detection} & Positive accuracy (\%) & 79.00 \\ & Negative accuracy (\%) & 88.00 \\ & Overall accuracy (\%) & 83.50 \\ & Latency (s) & 0.187 \\ \midrule \multirow{5}{*}{Temporal} & R1@0.3 (\%) & 48.50 \\ & R1@0.5 (\%) & 39.00 \\ & R1@0.7 (\%) & 31.00 \\ & mIoU (\%) & 38.73 \\ & Latency (s) & 0.967 \\ \bottomrule \end{tabular} \end{table}

As shown in Table~\ref{tab:ondevice_results}, the on-device AQA component achieves 79.00\% accuracy on positive detection questions and 88.00\% on negative detection questions, corresponding to an overall detection accuracy of 83.50\%. For temporal grounding, it achieves R1@0.3, R1@0.5, and R1@0.7 scores of 48.50\%, 39.00\%, and 31.00\%, respectively, with an mIoU of 38.73\%. The measured query-time latency is 0.187~s for detection questions and 0.967~s for temporal grounding questions. These measurements provide quantitative evidence that the downstream AQA component can execute on the target edge platform. However, the reported evaluation should not be interpreted as an end-to-end on-device benchmark of LA-RAG because AGM execution is excluded from the measurements and the on-device evaluation uses Qwen3-4B-2507 rather than the language model configuration used in the main experiments. The AGM deployment described above therefore demonstrates integration feasibility, while Table~\ref{tab:ondevice_results} specifically characterizes the on-device performance of the downstream structured reasoning stage.

\section{Summary Evaluation with an LLM Judge}
\label{app:llmasjudge}

We evaluate summary answers using GPT-4o as an LLM judge. For each example, the judge receives the question, the benchmark reference answer, and the system-generated summary, and assigns a single overall integer score from 1 to 5 based on content coverage, factual accuracy, completeness, and coherence. The rubric assigns a score of 5 to accurate, well-structured summaries covering all key reference information; 4 to summaries with minor omissions or presentation issues but no factual errors; 3 to partially complete summaries with missing important details or some inaccuracies; 2 to summaries with major omissions or substantial errors; and 1 to predominantly incorrect or irrelevant responses. Different wording is accepted when the meaning is preserved.

Evaluation uses Azure OpenAI Chat Completions with deployment \texttt{gpt4otest}, API version \texttt{2023-05-15}, temperature 0.1. Each successful evaluation uses one response, without averaging repeated judge ratings. API exceptions are retried for up to three total attempts, with a 10-second delay between attempts. The prompt requests JSON containing the overall score, qualitative content-coverage and factual-accuracy assessments, and a justification; no API-enforced JSON schema is used.

\paragraph{Judge input and prompt.}
The system message and summary user-prompt template are reproduced below.
The placeholders \texttt{\{question\}}, \texttt{\{ground\_truth\}},
and \texttt{\{predicted\}} are replaced by the question,
benchmark reference answer, and generated answer, respectively.
No full event table, audio, or additional context is supplied to the judge.
The reference and prediction correspond to the stored
\texttt{original\_answer} and \texttt{final\_answer} fields.

\paragraph{System message.}
\begin{Verbatim}[fontsize=\footnotesize,breaklines=true,breakanywhere=true]
You are an expert evaluator for audio analysis QA systems. Provide objective, consistent evaluations based on the rubric. Always return valid JSON.
\end{Verbatim}

\paragraph{Summary user-prompt template.}
\begin{Verbatim}[fontsize=\footnotesize,breaklines=true,breakanywhere=true]
You are evaluating system's answer to a SUMMARY question.

**Question:** {question}

**Ground Truth Answer:** {ground_truth}

**Predicted Answer:** {predicted}

**Evaluation Task:**
Compare the semantic content, completeness, and accuracy of the predicted summary against the ground truth.

**Scoring Rubric (1-5):**

5 - Excellent: Covers all key points from ground truth + accurate + well-structured
    - All major events/facts mentioned
    - Correct temporal/causal relationships
    - No significant omissions or hallucinations

4 - Good: Covers most key points + mostly accurate + minor issues
    - 1-2 minor omissions of secondary details
    - Slight verbosity or organizational issues
    - No factual errors

3 - Acceptable: Covers some key points + partially accurate
    - Missing 1-2 important details
    - Some factual inaccuracies or unclear statements
    - Core message is preserved

2 - Poor: Misses major points OR contains significant errors
    - Missing multiple important details
    - Contains factual errors or contradictions
    - Partially relevant but incomplete

1 - Inadequate: Mostly incorrect or irrelevant
    - Misses most key information
    - Contains major factual errors
    - Does not address the question properly

**Evaluation Criteria:**
- Content Coverage: Are all key events/facts included?
- Factual Accuracy: Are the details correct?
- Completeness: Is important context provided?
- Coherence: Is the summary well-organized and clear?

**Note:** Different wording is acceptable if the semantic meaning is preserved.

**Output Format (JSON):**
{
    "score": <1-5>,
    "content_coverage": "<high/medium/low>",
    "factual_accuracy": "<high/medium/low>",
    "reasoning": "<detailed explanation of score>"
}
\end{Verbatim}
```

\subsection{Human Validation of Summary Evaluation} \label{app:human_summary_audit}
\paragraph{Human agreement with GPT-4o evaluation.} Five human raters completed 100 rating assignments drawn from a stratified sample of 60 LA-RAG outputs, with each rater assigned 20 outputs. Raters received the question, reference answer, and generated answer and applied an overall 1-5 rubric adapted from the automated evaluator. Ratings flagged for insufficient reference information were excluded individually, while other available ratings for the same question were retained. The primary analysis contains 65 ratings over 45 questions: 20 questions with two retained ratings and 25 with one. GPT-4o scores show strong agreement with the mean available human rating, with Spearman's $\rho=0.71$ (95\% bootstrap CI: $[0.54,0.84]$), a mean absolute difference of 0.80 points, and a mean GPT-minus-human difference of $-0.56$ points. Agreement is stronger for event-specific summaries ($\rho=0.80$; Table~\ref{tab:human-gpt-summary}). On the 20 double-rated questions, human ordinal Krippendorff's $\alpha$ is 0.56, with 45\% exact agreement and 90\% agreement within one point. These results indicate that GPT-4o scores provide an informative approximation of human judgments, particularly for event-specific summaries.

\begin{table}[t]
\centering
\small
\begin{tabular}{lrrrr}
\toprule
Subset & $N$ & $\rho$ & MAD & Bias \\
\midrule
Overall        & 45 & 0.71 & 0.80 & $-0.56$ \\
Home-IoT       & 23 & 0.75 & 0.87 & $-0.70$ \\
Industrial-IoT & 22 & 0.58 & 0.73 & $-0.41$ \\
\midrule
Event-specific & 30 & 0.80 & 0.57 & $-0.33$ \\
Generic        & 15 & 0.61 & 1.27 & $-1.00$ \\
\bottomrule
\end{tabular}
\caption{GPT-human agreement after excluding individual ratings flagged
for insufficient reference information. $N$ is the number of questions;
$\rho$ is Spearman correlation with the mean available human rating.
MAD is mean absolute score difference, and bias is mean GPT minus human
score, both on the 1-5 scale. The dataset and summary-type rows are two
separate partitions of the overall set. These are agreement statistics,
not accuracy scores.}
\label{tab:human-gpt-summary}
\end{table}

\section{Grounding Mode Implementation Details} \label{app:grounding_details} \subsection{Offline Grounding Pipeline} Offline grounding consists of five stages. First, AGM processes the audio stream using overlapping sliding windows. It normalizes the detected events and stores them in an SQL database as structured records. Second, the query-rephrasing module uses the conversation history to resolve references in follow-up queries. For example, a query such as ``How many times did it happen?'' is rewritten to include the previously referenced event and time range. Third, the time-resolution module maps expressions such as ``before noon,'' ``last night,'' and ``during the morning shift'' to absolute timestamps. Common expressions are handled using rules, while an LLM is used as a fallback for ambiguous expressions. Fourth, a hybrid classifier assigns the resolved query to one of four intents: \textit{detection}, \textit{counting}, \textit{summary}, or \textit{anomaly}. The classifier combines keyword rules and embedding similarity; queries below the similarity threshold are assigned the summary intent. Finally, LA-RAG retrieves events from the SQL store using the resolved temporal constraints, structured filters, and embedding similarity. The retrieved event records are inserted into an intent-specific prompt and passed to the language model for answer generation. This design moves acoustic processing outside the query path and retains explicit timestamped evidence for each generated answer. \subsection{Inference-Time Grounding Pipeline} 
\begin{figure*}[!ht] \centering \includegraphics[width=0.8\linewidth]{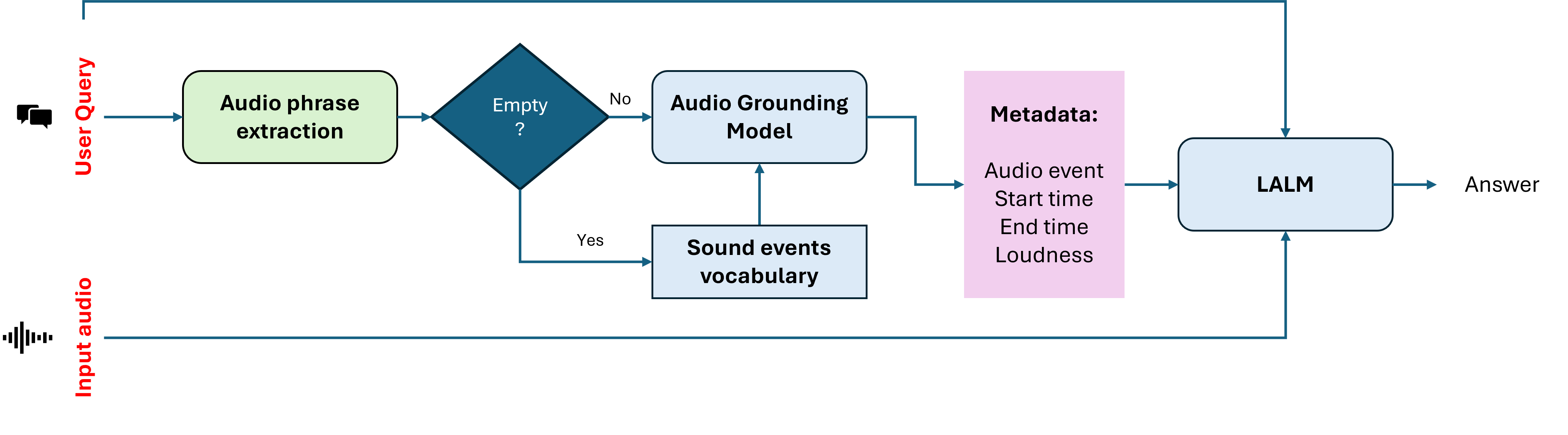} \caption{Query-conditioned grounding during inference.} \label{fig:inference_time_agm} \end{figure*}

Inference-time grounding modifies the AGM processing stage to perform query-conditioned grounding. After query rephrasing, a rule-based regular-expression pipeline removes common question prefixes and stop words to extract the acoustic phrase. For example, ``Is there a door knock in the audio?'' produces the acoustic phrase ``door knock.'' For compound queries containing connectors such as ``while,'' ``alongside,'' or ``as,'' the extracted phrase is divided into event-specific sub-phrases, and each sub-phrase is passed independently to AGM. The extracted phrase is supplied to AGM as a candidate event tag. AGM processes the input audio and returns timestamps and confidence scores for detections corresponding to that event. This avoids broad-vocabulary detection when the query already specifies the target sound. If phrase extraction fails, the system evaluates a fallback vocabulary of 75 acoustic tags derived from the AudioSet ontology following~\cite{kim2019audiocaps}. This vocabulary is used only for generic queries; query-specific grounding is not restricted to these classes. The detected event metadata are combined with the original query, inferred intent, and temporal constraints. Depending on the evaluated configuration, this information is passed either to a text-only LLM or to a LALM that also receives the raw audio. Figure~\ref{fig:inference_time_agm} illustrates the pipeline. 

\end{document}